\documentclass[twocolumn,floats,showpacs,prd,amssymb,floatfix,nofootinbib,balancelastpage,superscriptaddress,amsmath]{revtex4}
\usepackage{psfig}
\usepackage{epsfig}

\newcommand{\stau}{\widetilde{\tau_1}}
\newcommand{\neut}{\chi^0_1}

                              \newlength{\strikewidth}
                              \newlength{\strikelength}
\begin{document}

\title{A Running Spectral Index in Supersymmetric 
Dark-Matter \\ Models with Quasi-Stable Charged Particles}

\author{Stefano Profumo}
\email{profumo@sissa.it}
\affiliation{SISSA/ISAS, via Beirut 2-4, 34013 Trieste, Italy}
\author{Kris Sigurdson}
\email{ksigurds@tapir.caltech.edu}
\affiliation{California Institute of Technology, Mail Code 130-33, Pasadena, CA
91125}
\author{Piero Ullio}
\email{ullio@sissa.it}
\affiliation{SISSA/ISAS, via Beirut 2-4, 34013 Trieste, Italy}
\author{Marc Kamionkowski}
\email{kamion@tapir.caltech.edu}
\affiliation{California Institute of Technology, Mail Code 130-33, Pasadena, CA
91125}


\begin{abstract}

We show that charged-particles decaying in the early Universe can induce a 
scale-dependent or `running' spectral index in the small-scale linear and nonlinear matter 
power spectrum and discuss examples of this effect in minimal supersymmetric 
models in which the lightest neutralino is a viable cold-dark-matter candidate.
We find configurations in which the neutralino relic density is set by coannihilations 
with a long-lived stau, and the late decay of staus partially suppresses the linear 
matter power spectrum.  Nonlinear evolution on small scales then causes the
modified linear power spectrum to evolve to a nonlinear power spectrum similar (but different 
in detail) to models parametrized by a constant running $\alpha_{s}=d n_{s}/d{\rm ln} k$ 
by redshifts of 2 to 4.  Thus, Lyman-$\alpha$ forest observations, which probe the matter power spectrum at these redshifts, might not discriminate between the two effects.  However, a measurement of the angular power spectrum of primordial 21-cm 
radiation from redshift $z \approx 30$--$200$ might distinguish between this 
charged-decay model and a primordial running spectral index. The direct production 
of a long-lived charged particle at future colliders is a dramatic prediction of this model.

\end{abstract}




\maketitle

\section{Introduction}

While recent cosmological observations provide convincing evidence that nonbaryonic dark matter exists \cite{CMBconstraints}, we do not know the detailed particle properties of the dark matter, nor the particle spectrum of the dark sector.  There has been considerable phenomenological effort towards placing model-independent limits on the possible interactions of the lightest dark-matter particle (LDP)\footnote{In supersymmetric models the LDP is the lightest supersymmetric particle (LSP), but we adopt this more general notation unless we are speaking about a specific supersymmetric model.} in an attempt to try and identify candidates within detailed particle theories or rule out particular candidate theories.  For instance, models with stable charged dark matter have been ruled out \cite{Gould:1989gw}, while significant constraints have been made to dark-matter models with strong interactions \cite{Starkman:1990nj}, self-interactions \cite{Carlson:1992,Spergel:1999mh}, and a millicharge \cite{Davidson:2000hf,Dubovsky:2003yn}.  Recently, it was shown that a neutral dark-matter particle with a relatively large electric or magnetic dipole moment remains a phenomenologically viable candidate \cite{Sigurdson:2004zp}. 

Concurrent with this phenomenological effort, theorists have
taken to considering physics beyond the standard model in search
of a consistent framework for viable dark-matter candidates.
The leading candidates, those that produce the correct relic
abundance and appear in minimal extensions of the standard model
(often for independent reasons), are the axion~\cite{axionreviews} and
weakly interacting massive particles (WIMPs), such as the neutralino, 
the lightest mass eigenstate from the superposition of the supersymmetric 
partners of the $U(1)$ and $SU(2)$ neutral gauge bosons and of the 
neutral Higgs bosons~\cite{Jungman:1995df,larsrev}.  
However, other viable candidates have also
been considered recently, such as gravitinos or Kaluza-Klein
gravitons produced through the late decay of WIMPs
\cite{Feng:2003xh,Feng:2004zu}.  These latter candidates are an
interesting possibility because the constraints to the
interactions of the LDP do not apply to the next-to-lightest
dark-matter particle (NLDP), and the decay of the NLDP to the
LDP at early times may produce interesting cosmological effects;
for instance, the reprocessing of the light-element abundances
formed during big-bang nucleosynthesis \cite{Feng:2003uy,karsten}, or if
the NLDP is charged, the suppression of the matter power
spectrum on small scales and thus a reduction in the expected
number of dwarf galaxies \cite{Sigurdson:2003vy}.

In this paper we describe another effect charged NLDPs could
have on the matter power spectrum.  If all of the present-day
dark matter is produced through the late decay of charged NLDPs,
then, as discussed in Ref.~\cite{Sigurdson:2003vy}, the effect
is to essentially cut off the matter power spectrum on scales
that enter the horizon before the NLDP decays.  However, if only
a fraction $f_{\phi}$ of the present-day dark matter is produced through 
the late decay of charged NLDPs, the matter power spectrum is 
suppressed on small scales only by a factor $(1-f_{\phi})^2$.  This induces a scale-dependent spectral index for wavenumbers that enter the horizon when the age of the Universe is equal to the lifetime of the charged particles.
What we show below is that, for certain combinations of
$f_{\phi}$ and of the lifetime of the charged particle $\tau$, 
this suppression modifies the nonlinear power spectrum in a way similar 
(but different in detail) to the effect of a constant $\alpha_{s}
\equiv d n_{s}/d{\rm ln}k \neq 0$.  Although these effects are different,
constraints based on observations that probe the nonlinear power
spectrum at redshifts of 2 to 4, such as measurements of the
Lyman-$\alpha$ forest, might confuse a running index with the effect we describe here even if parametrized in terms of a constant $\alpha_{s}$.  This
has significant implications for the interpretation of the
detection of a large running of the spectral index as a
constraint on simple single-field inflationary models.  The
detection of a unexpectedly large spectral running in future observations could
instead be revealing properties of the dark-matter particle
spectrum in conjunction with a more conventional model of
inflation.  We note that the Sloan Digital Sky Survey Lyman-$\alpha$ data \cite{Seljak:2004xh} has 
significantly improved the limits to constant-$\alpha_{s}$ models compared to 
previous measurements alone \cite{CMBconstraints,Seljak:2003jg,Tegmark:2003ud}. A detailed study of the ($\tau$,$f_{\phi}$) parameter space using these and other cosmological data would also provide interesting limits to the models we discuss here.  

While, even with future Lyman-$\alpha$ data, it may be difficult to discriminate the effect of a constant running of the spectral index from a scale-dependent spectral index due to a charged NLDP, other observations may nevertheless discriminate between the two scenarios.  Future measurements of the power
spectrum of neutral hydrogen through the 21cm-line might probe the
linear matter power spectrum in exquisite detail over the
redshift range $z \approx 30 - 200$ at comoving scales less than
1 Mpc and perhaps as small as $0.01$ Mpc \cite{Loeb:2003ya};
such a measurement could distinguish between the charged-particle decay scenario we describe here and other modifications to the primoridal power spectrum.
If, as in some models we discuss below, the mass of these particles is in reach of future particle colliders the signature of this scenario would be spectacular
and unmistakable---the production of very long-lived charged
particles that slowly decay to stable dark matter.

Although we describe the cosmological side of our calculations
in a model-independent manner, remarkably, there are
configurations in the minimal supersymmetric extension 
of the standard model  (MSSM) with the right properties for the effect 
we discuss here. In particular, we find that if the LSP is a neutralino 
quasi-degenerate in mass with the lightest stau, we can naturally 
obtain, at the same time, LDPs providing the correct dark matter abundance 
$\Omega_{\chi}h^2 = 0.113$ \cite{CMBconstraints} and NLDPs 
with the long lifetimes and the sizable densities in the early Universe 
needed in the proposed scenario. Such configurations arise even
in minimal supersymmetric schemes, such as the minimal
supergravity (mSUGRA) scenario~\cite{msugra} and the minimal anomaly-mediated 
supersymmetry-breaking (mAMSB) model~\cite{mamsb}. This 
implies that a detailed study of the ($\tau$,$f_{\phi}$) parameter space 
using current and future cosmological data may constrain regions 
of the MSSM parameter space that are otherwise viable.  Furthermore, we are able to make quantitative statements about
testing the scenario we propose in future particle colliders or dark matter 
detection experiments.

The paper is organized as follows: We first review in
Section~\ref{sec:charged} how the standard calculation of
linear perturbations in an expanding universe must be modified
to account for the effects of a decaying charged species, calculate the linear matter power spectrum and discuss the constraints to this model from big bang nucleosynthesis (BBN) and the spectrum of the cosmic microwave background (CMB).  In
Section~\ref{sec:nonlinear} we briefly discuss how we estimate
the nonlinear power spectrum from the linear power spectrum and
present several examples.  In Section~\ref{sec:21cm} we discuss
how measurements of the angular power spectra of the primordial
21-cm radiation can be used to distinguish this effect from other modifications 
to  the primordial power spectrum.  In Section~\ref{sec:partmodel} we
describe how this scenario can be embedded in a particle physics model, 
concluding that the most appealing scheme is one
where long lifetimes are obtained by considering nearly
degenerate LDP and NLDP masses.
In Section~\ref{sec:lifetimes} we compute the lifetimes of charged 
next-to-lightest supersymmetric particles (NLSPs) decaying into neutralino  
LSPs, and show that, in the MSSM, the role of a NLDP with a long lifetime can be 
played by a stau only. In Section~\ref{sec:fraction} we estimate what 
fraction of charged to neutral dark matter is expected in this 
case, while in Section~\ref{sec:msugra} we  describe how consistent
realizations of this scenario can be
found within the parameter space of mSUGRA and mAMSB.  Finally, in
Section~\ref{sec:colliders} we discuss the expected signatures
of this scenario at future particle colliders, such as the
large hadron collider (LHC), and prospects for detection in experiments
searching for WIMP dark matter.  We conclude with a brief summary
of our results in Section~\ref{sec:conclusion}.

\section{Charged-Particle Decay}
\label{sec:charged}

In this section we discuss how the decay of a charged particle $\phi$ (the NLDP) to a neutral particle $\chi$ (the LDP) results in the suppression of the linear matter power spectrum on small scales.

As the $\phi$ particles decay to $\chi$ particles, their comoving energy 
density decays exponentially as 
\begin{align}
\rho_{\phi}a^{3}=m_{\phi} n_{\phi_0}e^{-t/\tau} \, ,
\label{eqn:rhophi}
\end{align}
increasing the comoving energy density of $\chi$ particles as
\begin{align}
\rho_{\chi}a^{3}=m_{\chi}n_{\chi_0}(1-f_{\phi} e^{-t/\tau}) \, .
\label{eqn:rhochi}
\end{align}
Here $n_{\chi_0}=\Omega_{\chi}\rho_{crit}/m_{\chi}$ is the
comoving number density of dark matter, and $n_{\phi_0}=f_{\phi}
n_{\chi_0}$ is the comoving number density of dark matter
produced through the decay of $\phi$ particles, $a$ is the scale
factor, and $t$ is the cosmic time.

Since the $\phi$ particles are charged, they are tightly
coupled to the ordinary baryons (the protons, helium nuclei, and
electrons) through Coulomb scattering.  It is therefore possible
describe the combined $\phi$ and baryon fluids as a generalized
baryon-like component $\beta$ as far as perturbation dynamics is
concerned. We thus denote by $\rho_{\beta}=\rho_{b}+\rho_{\phi}$
the total charged-particle energy density at any given time.  At
late times, after nearly all $\phi$ particles have decayed, $
\rho_{\beta} \simeq \rho_{b}$.

The relevant species whose perturbation dynamics are modified
from the standard case are the
stable dark matter (subscript $\chi$), the charged species
(subscript $\beta$), and the photons (subscript $\gamma$).  By imposing covariant conservation of the
total stress-energy tensor, accounting for the Compton
scattering between the electrons and the photons, and
linearizing about a Friedmann-Robertson-Walker (FRW) Universe we
arrive at the equations describing the evolution of linear fluid
perturbations of these components in an expanding Universe.
\begin{figure}
\centerline{\psfig{file=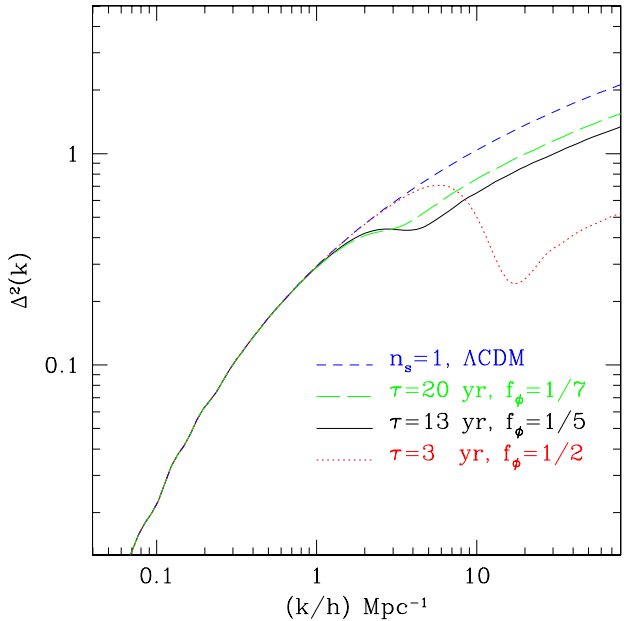,width=3.7in,angle=0}}
\caption{Shown is $\Delta^2(k)=k^3P(k)/2\pi^2$, the dimensionless matter power spectrum per logarithmic interval for the canonical $n_s=1$ $\Lambda$CDM model (dashed), for a model with $\tau=20~{\rm yr}$ and $f_{\phi}=1/7$ (long-dashed), for a model with $\tau=13~{\rm yr}$ and $f_{\phi}=1/5$ (solid), and for a model with $\tau=3~{\rm yr}$ and $f_{\phi}=1/2$ (dotted).  Modes that enter the horizon be before the $\phi$ particles decay are suppressed by a factor of $(1-f_{\phi})^2$.}
\label{fig:linear}
\end{figure}
In the synchronous gauge, for the `$\beta$' component, the perturbation evolution equations are
\begin{align}
\dot{\delta}_{\beta}  = - \theta_{\beta}-\frac{1}{2}\dot{h} \, ,
\label{eqn:delta_beta_2}
\end{align}
and
\begin{align}
\dot{\theta}_{\beta} = -\frac{\dot{a}}{a}\theta_{\beta} &+ c_{s}^2k^2\delta_{\beta} + \frac{4\rho_{\gamma}}{3\rho_{\beta}} a n_{e} \sigma_{T} (\theta_{\gamma}-\theta_{\beta}) \, .
\label{eqn:theta_beta_2}
\end{align}
Here and in what follows, $\delta_{X}={\delta\rho}_{X}/\rho_{X}$
is the fractional overdensity and $\theta_{X} = ikV_{X}$ is the
divergence of the bulk velocity in Fourier space of a given
species $X$.  An overdot represents a derivative with respect to
the conformal time.  The number density of electrons is $n_{e}$,
while $\sigma_{T}$ is the Thomson cross section.
Because the $\phi$ particles and the baryons share a common bulk
velocity and overdensity,\footnote{This assumes adiabatic initial
conditions. Note that the perturbation $S_{\phi\beta}
\equiv \delta_{\phi}-\delta_{\beta}$ will generally evolve away
from zero in an arbitrary gauge, even when starting with the
adiabatic initial condition $S_{\phi\beta}(0)=0$, due to gradients in the proper time.  It is a
special simplifying property of the synchronous gauge that
$S_{\phi\beta}$ = 0 for all time for adiabatic initial
conditions.} these equations are identical to the standard
perturbation equations for the baryons  with the replacement $b
\rightarrow \beta$ (see, for example, Ref.~\cite{Ma:1995ey}). 
For the dark matter we find that
\begin{align}
\dot{\delta}_{\chi}  = - \theta_{\chi}-\frac{1}{2}\dot{h} + \lambda_m \frac{\rho_{\phi}}{\rho_{\chi}}\frac{a}{\tau}(\delta_{\beta}-\delta_{\chi}) \, ,
\label{eqn:deltadot_d_2}
\end{align}
and
\begin{align}
\dot{\theta}_{\chi} = -\frac{\dot{a}}{a}\theta_{\chi} + \lambda_m \frac{\rho_{\phi}}{\rho_{\chi}}\frac{a}{\tau}(\theta_{\beta}-\theta_{\chi}) \, ,
\label{eqn:thetadot_d_2}
\end{align}
where $\lambda_m \equiv m_{\chi}/m_{\phi} = (1+\Delta m/m_{\chi})^{-1}$.
The modifications to the photon perturbation evolution are
negligibly small because $\phi$ decays during the
radiation-dominated epoch when $\rho_{\phi} \ll \rho_{\gamma}$
and, as discussed below, for viable models $\lambda_m \simeq 1$ to
prevent unreasonably large spectral distortions to the CMB. 

Combining these equations with the (unmodified) equations for
the neutrino perturbations we can solve for the linear power
spectrum of matter fluctuations in this model.  We have solved
these equations using a modified version of {\tt cmbfast}
\cite{Seljak:1996is}.  In Fig.~\ref{fig:linear} we show the
linear matter power spectrum in this model for several values of
the $\phi$ lifetime $\tau$ and fraction $f_{\phi}$.  As shown in
this Figure, the small-scale density modes that enter the
horizon before the $\phi$ particles decay (when the age of the
Universe is less than $\tau$) are suppressed relative to the
standard case by a factor of $(1-f_{\phi})^2$.

Since the decaying particles are charged, the production of the LDP will always be accompanied by an electromagnetic cascade.  The latter could in principle reprocess the light elements produced during BBN, or induce unreasonably large spectral distortions to the CMB.  We show here that in fact these effects are small for the models discussed in this paper.

The energy density released by the decay of $\phi$ particles can be parametrized as
\begin{equation}
\zeta_{EM} = \varepsilon_{EM} f_{\phi} Y_{\chi} \, ,
\end{equation}
where $\varepsilon_{EM}$ is the average electromagnetic energy released in a $\phi$ decay and $Y_{\chi} \equiv n_{\chi}/n_{\gamma}$ is the dark-matter to photon ratio.  In the specific models we discuss below, $\varepsilon_{EM} \approx \Delta m/3$, and 
\begin{equation}
Y_{\chi} = \frac{\Omega_{\chi} \rho_c}{m_{\chi} n_{\gamma}} = 3 \times 10^{-12} \left(\frac{{\rm TeV}}{m_{\chi}}\right)\left( \frac{\Omega_{\chi}}{0.23}\right) \, .
\end{equation}
This yields
\begin{equation}
\zeta_{EM} \approx f_{\phi} \frac{\Delta m}{m_{\chi}} \left(\frac{\Omega_{\chi}}{0.23}\right)~{\rm eV} \, .
\end{equation}
In the models we discuss below, $\Delta m/m_{\chi} \sim 10^{-4}$, and $f_{\phi} \leq 1/2$, giving $\zeta_{EM} \lesssim 5 \times 10^{-5}$~eV.  The limit derived from too much reprocessing of the BBN light element abundances is $\zeta_{EM} \lesssim 3.8~\tau_{\rm yr}^{1/4} \times 10^{-3}$~eV \cite{Cyburt:2002uv,jedamzik} where $\tau_{\rm yr}=\tau/(1 ~{{\rm yr}})$, so we are safely below this bound.

For $\tau_{\rm yr} \lesssim 300$, electromagnetic energy injection will result in a chemical-potential distortion to the CMB of \cite{Hu:1993gc}
\begin{align}
\mu = 4.5~\tau_{\rm yr}^{1/2} \times 10^{-3} \left(\frac{\zeta_{EM}}{\rm eV}\right) e^{-0.128\tau_{\rm yr}^{-5/4}} \,
\end{align}
and so we expect $\mu \lesssim 2.0 \times 10^{-7}$~---~$2.3 \times 10^{-6}$ for lifetimes between $\tau_{\rm yr} = 1$~---~$100$, below the current limit of $\mu < 9 \times 10^{-5}$ \cite{Fixsen:1996nj}.

\section{The Nonlinear Power Spectrum}
\label{sec:nonlinear}

As density perturbations grow under the influence of gravity, linear evolution ceases to describe their growth and nonlinear effects must be taken into account.  On large scales, where density perturbations have had insufficient time to become nonlinear, the linear matter power spectrum describes the statistics of density fluctuations.  However, on small scales the full nonlinear matter power spectrum is required.

\begin{figure}
\centerline{\psfig{file=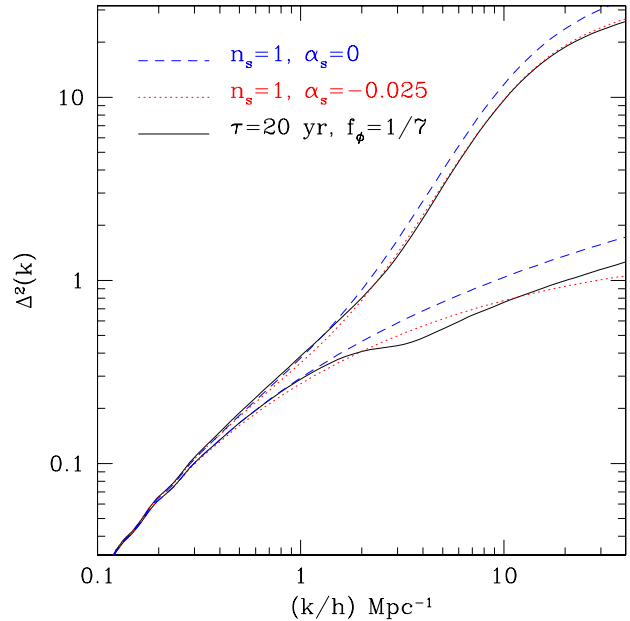,width=3.7in,angle=0}}
\caption{Shown at a redshift $z=4$ are
$\Delta^2(k)=k^3P(k)/2\pi^2$, the nonlinear (upper curves) and
linear (lower curves) dimensionless matter power spectra per
logarithmic interval for the canonical $n_s=1$ $\Lambda$CDM
model (dashed), for an $n_s=1$ model with $\tau=20~{\rm yr}$ and
$f_{\phi}=1/7$ (solid), and for a running-index model with
$n_s=1.00$ and $\alpha_s=-0.025$ (dotted).  Although the linear
power spectra differ significantly in these latter two models, nonlinear
evolution causes them to have nearly degenerate nonlinear power
spectra for $k/h \gtrsim 1.5~{\rm Mpc}^{-1}$.}
\label{fig:nonlinear_t20}
\end{figure}

\begin{figure}
\centerline{\psfig{file=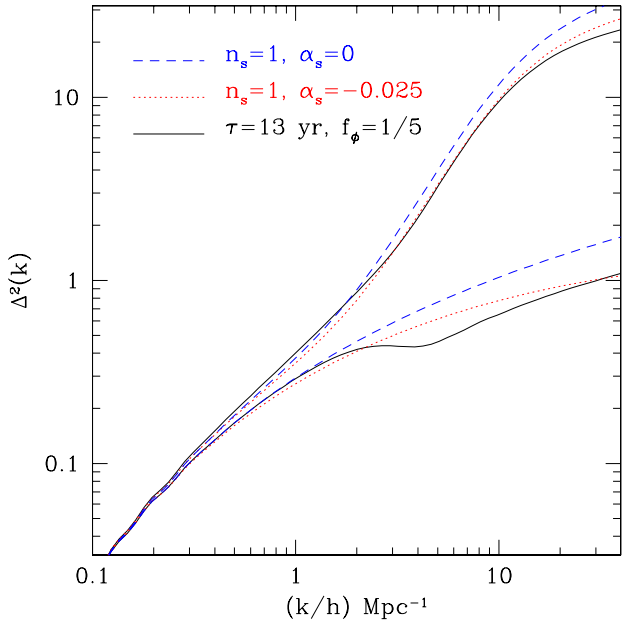,width=3.7in,angle=0}}
\caption{Shown at a redshift $z=4$ are
$\Delta^2(k)=k^3P(k)/2\pi^2$, the nonlinear (upper curves) and
linear (lower curves) dimensionless matter power spectra per
logarithmic interval for the canonical $n_s=1$ $\Lambda$CDM
model (dashed), for an $n_s=1$ model with $\tau=13~{\rm yr}$ and
$f_{\phi}=1/5$ (solid), and for a running-index model with
$n_s=1.00$ and $\alpha_s=-0.025$ (dotted).  Nonlinear evolution
causes these two models to have overlapping nonlinear power
spectra for $ 2~{\rm Mpc}^{-1} \lesssim k/h \lesssim 10~{\rm Mpc}^{-1}$.}
\label{fig:nonlinear_t13}
\end{figure}

\begin{figure}
\centerline{\psfig{file=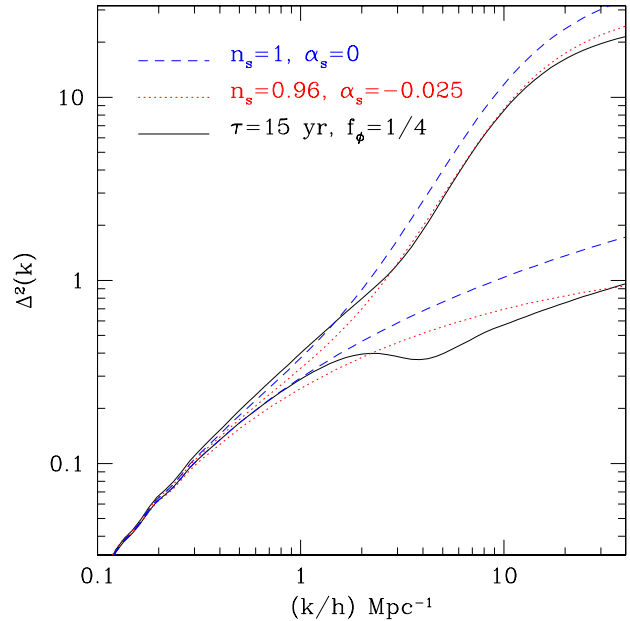,width=3.7in,angle=0}}
\caption{Shown at a redshift $z=4$ are
$\Delta^2(k)=k^3P(k)/2\pi^2$, the nonlinear (upper curves) and
linear (lower curves) dimensionless matter power spectra per
logarithmic interval for the canonical $n_s=1$ $\Lambda$CDM
model (dashed), for an $n_s=1$ model with $\tau=15~{\rm yr}$ and
$f_{\phi}=1/4$ (solid), and for a running-index model with
$n_s=0.96$ and $\alpha_s=-0.025$ (dotted).  Nonlinear evolution
causes the charged-decay model to match the canonical
$\Lambda$CDM model for $k/h \lesssim 1.5~{\rm Mpc}^{-1}$ and the
running-index model for $k/h \gtrsim 1.5~{\rm Mpc}^{-1}$.}
\label{fig:nonlinear_t15}
\end{figure}

\begin{figure}
\centerline{\hbox{\psfig{file=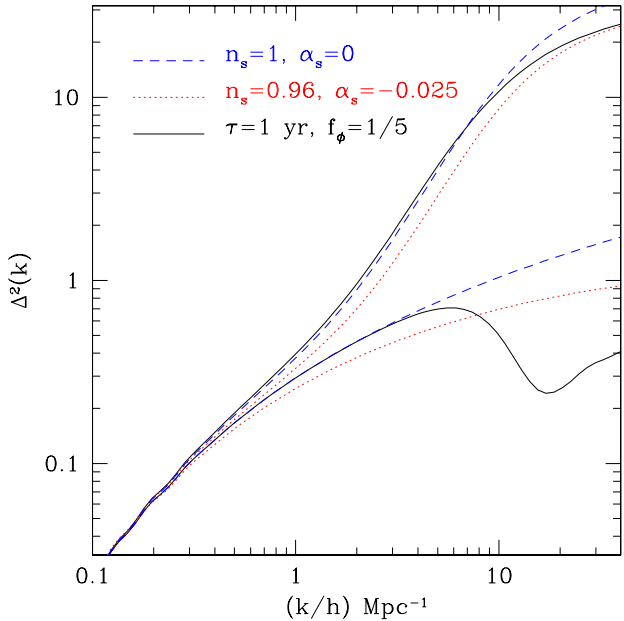,width=3.7in,angle=0}}}
\caption{Shown at a redshift $z=4$ are
$\Delta^2(k)=k^3P(k)/2\pi^2$, the nonlinear (upper curves) and
linear (lower curves) dimensionless matter power spectra per
logarithmic interval for the canonical $n_s=1$ $\Lambda$CDM
model (dashed), for an $n_s=1$ model with $\tau=1~{\rm yr}$ and
$f_{\phi}=1/2$ (solid), and for a running-index model with
$n_s=0.96$ and $\alpha_s=-0.025$ (dotted).  Despite the drastic change in the linear power spectrum nonlinear evolution
causes the charged-decay model to match the canonical
$\Lambda$CDM model for $k/h \lesssim 8~{\rm Mpc}^{-1}$ and the
running-index model for $k/h \gtrsim 8~{\rm Mpc}^{-1}$.}
\label{fig:nonlinear_t1}
\end{figure}

\begin{figure}
\centerline{\epsfig{file=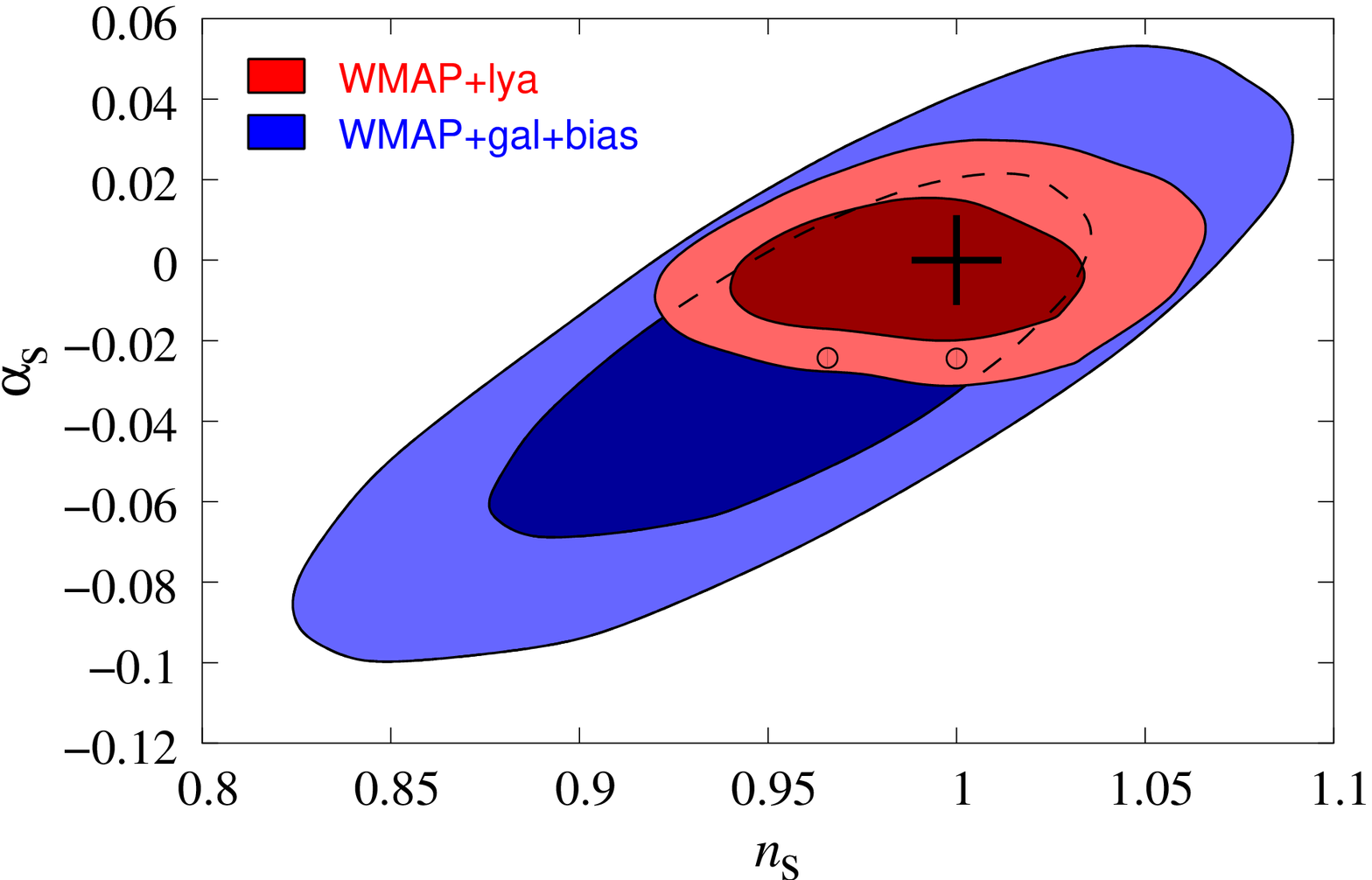,width=3.7in,angle=0,bbllx=0,bblly=13,bburx=555,bbury=358}}
\caption{We reproduce here (with permission) Fig.~3 from Ref.~\cite{Seljak:2004xh} which illustrates the current constraints on the parameter $\alpha_{s}$.  The charged-decay models produce changes to the nonlinear power spectrum similar to constant $\alpha_{s} \neq 0 $ models within the WMAP+Lya contour shown here and the examples we show in Figs.~\ref{fig:nonlinear_t20}--\ref{fig:nonlinear_t1} are denoted by the black circles (ours).  However, the charged-decay models leave the CMB angular power spectra unaltered (corresponding to models with $\alpha_{s} = 0$ on a line passing through the black cross). Matching to allowed constant-$\alpha_{s}$ models is conservative in this respect, and we thus expect the charged-decay models we have discussed in Figs.~\ref{fig:nonlinear_t20}--\ref{fig:nonlinear_t1} to be consistent with current data.  }
\label{fig:sdss_limits}
\end{figure}

In order to calculate the nonlinear power spectrum for a given
model we have used the recently devised {\tt halofit} method
\cite{Smith:2002dz}.  This method uses higher-order perturbation
theory in conjunction with the halo model of large-scale
structure to determine the nonlinear power spectrum given a
linear power spectrum.  It has been shown to accurately
reproduce the nonlinear power spectra of standard N-body
simulations and, unlike the earlier mappings such as the Peacock
and Dodds formula \cite{Peacock:1996ci}, it is applicable in
cases (like we consider here) when $\Delta^2(k)=k^3 P(k)/2\pi^2$
is not a monotonic function.  In particular we have checked that
it approximately reproduces the shapes of the nonlinear power
spectra determined in Ref.~\cite{White:2000sy} through N-body
simulations in models where the linear power spectrum is
completely cut off on small scales.  As we are discussing here
less drastic alterations to the linear power spectrum, we believe 
the {\tt halofit} procedure provides an estimate of
the nonlinear power spectrum adequate for illustrating the effect we describe in this paper.  Any detailed study would require a full N-body simulation.

In Figs.~\ref{fig:nonlinear_t20}--\ref{fig:nonlinear_t1} we show
both the linear and nonlinear matter power spectra at redshift $z=4$ (data from measurements the Lyman-$\alpha$ forest probe redshifts 2--4 at wavenumbers $k/h \sim 0.1$--$10~{\rm Mpc}^{-1}$) for 
a charged-decay model, and for a model with a running spectral
index.    Although these models have different linear power
spectra, nonlinear gravitational evolution causes these models
to have nearly identical nonlinear power spectra over
interesting ranges of wavenumbers.  The lifetimes shown were chosen because they produce effects on scales that can be probed by measurements of the Lyman-$\alpha$ forest, while the values $f_\phi=(1/2,1/4,1/5,1/7)$ were chosen because they arise in the supersymmetric models we discuss below.

In Fig.~\ref{fig:sdss_limits} we show the current constraints on the ($n_{s}$,$\alpha_{s}$) parameter space from WMAP and SDSS Lyman-$\alpha$ forest data.   The constant-$\alpha_{s}$ models we have compared our charged-decay model to lie within the WMAP+lya contours shown in the Figure. Since, in each case, the charged-decay model tends to interpolate between the standard $\Lambda$CDM model on larger 
scales (in particular on the scales probed by the CMB) and a constant-$\alpha_{s}$ running-index model on smaller scales, and 
both of them are allowed by these data, we expect the charged-decay  models we have
considered in Figs.~\ref{fig:nonlinear_t20}--\ref{fig:nonlinear_t1} to be consistent 
with current data as well.  We leave for future work the task of using a combined analysis of Lyman-$\alpha$ forest and other cosmological measurements to put limits directly on the ($f_{\phi}$,$\tau$) parameter space (and thus on the parameter space of the MSSM models we discuss below).

\section{The 21~cm Power Spectrum}
\label{sec:21cm}

After recombination and the formation of neutral hydrogen, the
gas in the Universe cools with respect to the CMB temperature
$T_{\rm CMB}$ starting at at a redshift $z \sim 200$.  The spin
temperature  $T_{s}$ of the gas, which measures the relative
populations of the hyperfine levels of the ground state of
hydrogen separated by the 21-cm spin-flip transition, remains
collisionally coupled to the temperature $T_b$ of the
baryons until a redshift $z\sim 30$ when collisions
become inefficient and the spin temperature rises to $T_{\rm
CMB}$. There is thus a window between $z \sim 30$--$200$ in
which neutral hydrogen absorbs the CMB at a wavelength of
$21~{\rm cm}$.  It has recently been suggested that the angular
fluctuations in the brightness temperature of the 21-cm
transition within this window may be measured with future
observations and used to constrain the matter power spectrum at
these very high redshifts \cite{Loeb:2003ya}.

At redshifts $z \gtrsim 30$, the matter power spectrum on the
scales $k/h \sim 1 - 100~{\rm Mpc}^{-1}$ of interest here are
still in the linear regime. As discussed in
Ref.~{\cite{Loeb:2003ya}}, due to the unprecedented wealth of
potential information contained in these $21$-cm measurements,
models with a running index or other small-scale modifications
of the matter power spectrum can in principle be distinguished
from each other.  Even
if the charged-particle lifetime is so small that no significant
modifications to the power spectrum occur on scales probed by
other cosmological observations (like the model shown in
Fig.~\ref{fig:nonlinear_t1}), 21-cm observations might
detect or constrain such effects on the linear matter power
spectrum.

\section{The long lived charged next-to-lightest dark-matter particle}
\label{sec:partmodel}

From the particle-physics point of view, the setup we have introduced 
may seem {\it ad hoc}. We need a pair of particles that share a
conserved quantum number and such that the lightest, the LDP, 
is neutral and stable, while the
other, the NLDP, is coupled to the photon and quasi-stable, in order to
significantly contribute to the cosmological energy density at
an intermediate stage in the structure-formation process.  
Such a picture requires three ingredients:
{\sl (i)} the relic abundance of the LDP must be compatible with
the CDM component;
{\sl (ii)} the abundance of the NLDP must be at the correct
level (namely, $\sim1/5$ the total dark-matter density); and
{\sl (iii)} the NLDP must have the proper lifetime (i.e.,
$\tau\sim10$ yr).

Let us start with the last requirement.  One way to get the required
lifetime is to introduce a framework with strongly-suppressed
couplings. One such possibility is, for instance, to assume that
the LDP is a stable super-weakly--interacting dark-matter
particle, such as a gravitino LSP in R-parity--conserving
supersymmetric theories or the
Kaluza-Klein first excitation $G^1$ of the graviton in the
universal extra-dimension scenario \cite{Servant:2002aq}. 
The NLDP can have non-zero electric charge, but at the same time
a super-weak decay rate into the LDP (with the latter being the
only allowed decay mode). In models of gauge-mediated
supersymmetry breaking, this might indeed be the case with a
stau NLSP decaying into a gravitino LSP.  In this specific
example, we have checked that, to retrieve the very long
lifetimes we  introduced in our discussion, we would need to
impose a small mass splitting between the NLSP and LSP, as well as
to raise the mass scale of the LSP up to about 100~TeV. This
value is most often considered uncomfortably large for a SUSY
setup, and it also makes thermal
production of LDPs or NLDPs unlikely, being near a scale at which the unitary
bound~\cite{Griest:1989wd} gets violated.  Without thermal production one would need
to invoke first a mechanism to wipe out the thermal components
and then provide a viable non-thermal production scheme that
fixes the right portion of LDPs versus NLDPs.  Finally, such
heavy and extremely weakly-interacting objects would evade any  
dark-matter detection experiment, and would certainly not be
produced at the forthcoming CERN Large Hadron Collider
(LHC). This would then be a scheme that satisfies the three
ingredients mentioned above, but that cannot be tested in any
other way apart from cosmological observations.

An alternative (and to us, more appealing) scenario is one
where long lifetimes are obtained by considering nearly
degenerate LDP and NLDP masses.  In this case, decay rates
become small, without suppressed couplings, simply 
because the phase space allowed in the decay process gets
sharply reduced.  Sizeable couplings imply that, in the early
Universe, the LDP and NLDP efficiently transform into each other
through scattering from background particles.  The small mass
splitting, in turn, guarantees that the thermal-equilibrium
number densities of the two species are comparable. To describe
the process of decoupling and find the thermal relic abundance
of these species, the number densities of the two have to be
traced  simultaneously, with the NLDP, being charged, playing
the major role. This phenomenon is usually dubbed as {\em
coannihilation}~\cite{Griest:1990kh} and has been studied
at length, being ubiquitous in many frameworks embedding
thermal-relic candidates for dark matter, including common
SUSY schemes.

We will show below that sufficiently long lifetimes may be
indeed obtained in the minimal supersymmetric standard model
(MSSM), when the role of the NLDP is played by a stau nearly
degenerate in mass with the lightest neutralino (with the former
being the stable LSP and the thermal-relic CDM candidate we will
focus on in the remainder of our discussion).  A setup of this kind
appears naturally, e.g., in minimal supergravity
(mSUGRA) \cite{Ellis:2003cw}.  This is the SUSY framework with
the smallest possible parameter space, defined by only four
continuous entries plus one sign, and hence also one of the most
severely constrained by the requirement that the neutralino
relic density matches the value from cosmological observations.
Neutralino-stau coannihilations determine one of the allowed regions, 
on the border with the region where the stau, which in the 
mSUGRA scheme is most often the lightest scalar SUSY particle, becomes 
lighter than the neutralino.  Although a stau-neutralino mass
degeneracy is not ``generic'' in such models, this scenario is
economical in that the requirements of a long NLDP lifetime and
of comparable LDP and NLDP relic abundances are both
consequences of the mass degeneracy.  In this sense, evidence
for a running spectral index or any of the other observational
features we discuss would simply help us sort out which
configuration (if any) Nature has chosen for SUSY dark matter.

\section{Lifetimes of charged NLSPs in the MSSM}
\label{sec:lifetimes}

We refer to a MSSM setup in which the lightest neutralino $\neut$
is the lightest SUSY particle.  The charged particles
that could play the role of the NLDP include: (1) scalar
quarks, (2) scalar charged leptons, and (3) charginos.  We now
discriminate among these cases by the number
of particles in the final states for the decay of NLSPs
to neutralino LSPs.

Scalar quarks and leptons have as their dominant decay mode a
prompt two-body final state; i.e., $\tilde S\rightarrow \neut
S$, where we have labeled $\tilde S$ the SUSY scalar partner of
the standard-model fermion $S$.  A typical decay width  
for this process is $\mathcal{O}$(1)~GeV, corresponding to a
lifetime $\mathcal{O}$$(10^{-24})$~s. This holds whenever this
final state is kinematically allowed; i.e., if $m_{\tilde
S}>m_S+m_{\neut}$.  If it is {\em kinematically forbidden}, there
are two possibilities: squarks may either decay through
CKM-suppressed flavor-changing processes or through four-body
decays. For instance, the stop decay may proceed through 
$\tilde t \rightarrow  \neut c$ or $\tilde t \rightarrow  \neut
b  f \bar f^\prime$.  On the other hand, within the same
minimal--flavor-violation framework, scalar leptons are not
allowed to decay in flavor-changing two-body final states, and
only the four-body decay option remains.  

The case for the chargino is different because this NLSP decay
has a three-body final state, either with two quarks bound in a
meson state --- i.e., $\chi^+_1 \rightarrow  \neut \pi^+$ --- or
with a leptonic three body channel --- i.e., $\chi^+_1\rightarrow
\neut l^+ \nu_l$.  The latter final state becomes dominant, in
particular, for electron-type leptons, $l=e$, if the mass
splitting between NLSP and LSP  becomes small.

We have listed all decay topologies as these are especially 
relevant when discussing the limit in which we force a reduction of the allowed
decay phase-space volume; i.e., the limit in which the NLSP and
LSP are quasi-degenerate in mass.  Here we can also safely
assume that the masses of the final-state particles, apart from
the neutralino, are much smaller than the mass of the decaying
particle.  We can then consider the limit of a particle of mass
$m_{\neut}+\Delta m$ decaying into a $\neut$ and $n-1$ massless
final states, and derive an analytical approximation to the
behavior of the final-state phase space ${\rm d}\phi^{(n)}$ 
and of the decay width $\Gamma^{(n)}$ as functions of $\Delta m$. 
In the case of two-body decays, the phase space reads
\begin{equation}
{\rm d}\phi^{(2)}=\frac{{\rm d}\Omega}{32\pi^2}\left(1-\left(\frac{m_{\neut}}{m_{\neut}+\Delta m}\right)^2\right)\propto \Delta m.
\end{equation} 
On the other hand, a recursive relation between ${\rm
d}\phi^{(n)}$ and ${\rm d}\phi^{(n-1)}$ based on the invariant
mass of couples of final states yields
\begin{eqnarray}
{\rm d}\phi^{(n)}(\Delta m)&\propto& {\rm d}\phi^{(n-1)}(\Delta m)\times \int^{\Delta m} {\rm d}\mu({\rm d}\phi^{(2)})(\mu) \nonumber\\
&\propto&  (\Delta m)^{2(n-2)+1}.
\end{eqnarray} 
The dependence of the decay width $\Gamma^{(n)}$ on $\Delta m$
must, however, take into account not only the phase-space
dependence, but also the behavior of the amplitude squared
${\cal M}^{(n)}$ of the processes as a function of $\Delta
m$. The occurrence of a massless final state yields, in the
amplitude squared, a factor that scales linearly with the momenta
circulating in the Feynman diagram. One therefore has the
further factor
\begin{equation}
{\cal M}^{(n)}\propto (\Delta m)^{n-1}.
\end{equation}
Finally, we have
\begin{equation}
\Gamma^{(n)}\propto {\cal M}^{(n)} \times {\rm d}\phi^{(n)} \propto (\Delta m)^{3n-4},\label{eq:scaling}
\end{equation}
i.e., the lifetime to decay to a two-body final state scales like
$\tau^{(2)}\propto (\Delta m)^{-2}$, while for a four-body decay
we have $\tau^{(4)}\propto (\Delta m)^{-8}$.
\begin{figure}
\centerline{\psfig{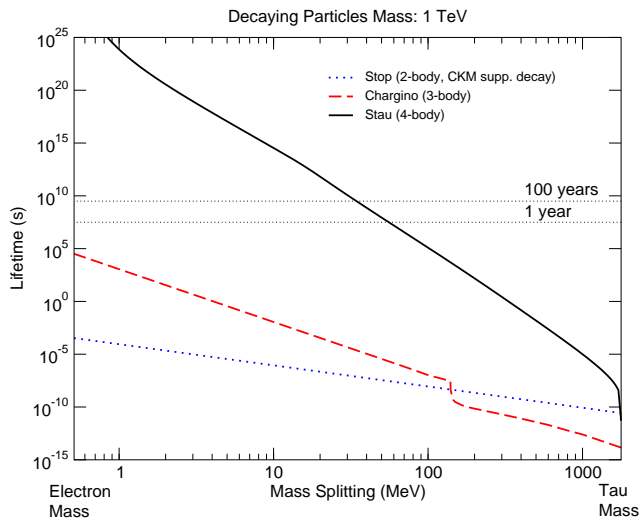}}
\caption{The lifetime of a 1 TeV stop (blue dotted line), a
chargino (red dashed line) and a stau (black solid line), as a
function of their mass splitting with the lightest SUSY
particle, the lightest neutralino.}
\label{fig:lifetimes}
\end{figure}

Reducing the NLSP mass splitting $\Delta m$ one may hope to
obtain ``cosmologically relevant'' NLSP lifetimes. For scalar
quarks, this is not quite the case, as two-body final states
always dominate, and even for amplitudes that are
CKM-suppressed, the scaling with $\Delta m$ is too shallow, and
the resulting lifetimes are rather short, even for very small
mass splittings.

The situation is slightly more favorable for charginos, whose
three-body decay width is approximately equal to
\begin{equation}
\frac{G_F^2}{(2\pi)^3}\frac{16}{15}(\Delta m)^5.
\end{equation} 
At tree level and in the limit of pure Higgsino-like or
Wino-like states, the lightest neutralino and lightest chargino
are perfectly degenerate in mass. However, when one takes into
account loop corrections to the masses, $\Delta m$ usually turns
out to be larger than a few tens of MeV; see, e.g.,~\cite{Pierce:1996zz,Cheng:1998hc}.
 This translates into an absolute upper limit on the chargino lifetime
of about $10^{-(2-3)}$ s.

Finally, in the case of sleptons $\tilde l$, the interesting regime is when 
$\Delta m<m_l$ and only four-body decays are allowed.  For
example, for the lightest stau $\stau$, the processes 
\begin{equation}
\stau\,\, \rightarrow \,\,  \neut \, \nu_\tau \, f\, \bar f^\prime,
\end{equation}
with
\begin{equation}\label{eq:fs}
(\bar f^\prime, f)=(\bar\nu_\mu,\mu),\,\, (\bar\nu_e,e),\,\, (\bar u,d),\,\, (\bar c,s),
\end{equation}
where, depending on $\Delta m$, only final states above the
kinematic threshold are included. In this case, $\Delta m$ can
be safely taken as a free parameter of the theory: in most
scenarios the gaugino mass parameter setting the neutralino
mass for bino-like neutralinos, and the scalar soft mass
parameter setting the stau mass, are usually assumed to be
independent; analogously, there are models in which the 
$\mu$ parameter setting the mass for a Higgsino-like neutralino,
and scalar soft mass parameter are unrelated. A similar picture applies
to the smuon, though lifetimes start to be enhanced at much
smaller mass splittings ($m_\mu$ instead of $m_\tau$), and it is
theoretically difficult to figure out a scenario in which the
lightest smuon is lighter than the lightest stau.

The scalings we have sketched are summarized in
Fig.~\ref{fig:lifetimes}, where we plot lifetimes for a stop
(CKM-suppressed), a chargino, and a stau as a function of
$\Delta m$. The decaying-particle masses have been set 
to 1 TeV, and the $\Delta m$ range is between the
electron and the tau mass. The lifetimes of the stop and of the
chargino have been computed with SDECAY
\cite{Muhlleitner:2003vg}. Some details on the computation of
the stau lifetime are given below. Notice that the scaling of
Eq.~(\ref{eq:scaling}) is accurately reproduced for all
cases. The bottom line is that indeed {\em the lightest stau can
play the role of the NLDP with a cosmologically relevant
lifetime}, and that the stau is the only particle in the MSSM
for which this can be guaranteed by adjusting only the LDP-NLDP
mass splitting.

\section{The Relative Abundance of the Charged NLSP}
\label{sec:fraction}

The next step is to determine the relic density of the NLSP and
LSP in the early stages of the evolution of the Universe.  As we
have already mentioned, we are going  to consider thermal
production.  We briefly review here how to compute
the evolution of number densities with coannihilation
\cite{Griest:1990kh,edsjogondolo}.

Consider a setup with $N$ supersymmetric particles $\chi_1$, $\chi_2$, 
... $\chi_N$, each with mass $m_i$ and number of internal degrees of
freedom $g_i$. The ordering is such that $m_1 \le m_2 \le \dots \le m_N$.
In the evolution equations, the processes that change the number density 
of SUSY particles $i$ are of three kinds:
\begin{equation}
\begin{array}{lcl}
{\rm (a)} \;\; 
\chi_i \; \chi_j \leftrightarrow X_a^f \, , & \hspace{1cm}&\forall \;j \, ,\\
{\rm (b)} \;\;
\chi_i \; X_b^i \leftrightarrow \chi_j \; X_b^f \, , & &\forall \;j\neq i \, ,\\
{\rm (c)} \;\;
\chi_j \leftrightarrow \chi_i \; X_c^f \, , & &\forall \; j>i \, ,
\end{array}
\label{eq:list}
\end{equation}
where $X_a$, $X_b^i$, $X_b^f$, and $X_c^f$ are (sets
of) standard-model (SM) particles.  In practice, the relevant
processes one should include are those for SM particles that are
in thermal equilibrium.  Assuming the distribution function for
each particle $k$ is the same as for the equilibrium  
distribution function,
\begin{equation}
f_k(E_k) \propto f_k^{\rm eq}(E_k) = \frac{1}{\exp(E_k/T) \pm 1}\;, 
\end{equation} 
and invoking the principle of detailed balance, the Boltzmann 
equation for the evolution of the number density of SUSY particle $i$,
$n_i = g_i/(2\pi)^3 \int d^3p f_i(E)$, normalized to
the entropy density of the Universe, $Y_i = n_i /s$, as a
function of the variable $x \equiv m_1/T$ (with $T$ the Universe
temperature; this is equivalent to describing the evolution in
time) is given by
\begin{widetext}
\begin{eqnarray}
\frac{x}{\hat{g}(x)\;Y_i^{\rm eq}}
\frac{dY_i}{dx} &=& 
- \sum_{j} \frac{\langle\sigma_{ij} v_{ij}\rangle \, n_j^{\rm eq}}{H}
\left(\frac{Y_i \, Y_j}{Y_i^{\rm eq} \, Y_j^{\rm eq}}-1\right)
 - \sum_{j\neq i}   
\frac{
\left[\sum_X \langle\sigma_{iX \rightarrow j} v_{iX \rightarrow j}\rangle\, 
n_X^{\rm eq} \right]}{H} 
\left(\frac{Y_i}{Y_i^{\rm eq}}-\frac{Y_j}{Y_j^{\rm eq}}\right) \\ \nonumber
&& + \sum_{j>i} \frac{\Gamma_{j \rightarrow i}}{H}
\left(\frac{Y_j}{Y_i^{\rm eq}}-\frac{Y_i \, Y_j^{\rm eq}}{(Y_i^{\rm eq})^2}\right) 
- \sum_{j<i} \frac{\Gamma_{i \rightarrow j}}{H}
\left(\frac{Y_i}{Y_i^{\rm eq}}-\frac{Y_j}{Y_j^{\rm eq}}\right) 
\;.
\label{eq:b3}
\end{eqnarray}
\end{widetext}
In this equation, analogously to $Y_k$, we have defined
$Y_k^{\rm eq} \equiv n_k^{\rm eq} /s$, the ratio of the equilibrium number density 
of species $k$ (at temperature $x$) to the entropy density.
On the left-hand side, we introduced $\hat{g}(x) \equiv
[1+T/(3\,g_{\rm eff})\,dg_{\rm eff}/dT]^{-1}$ with $g_{\rm
eff}(T)$ being the effective degrees of freedom in the entropy density.
The function $\hat{g}(x)$ is close to 1 except for temperatures
at which a background particle becomes nonrelativistic. On the
right-hand side, the last two terms contain factors in
$\Gamma_{k \rightarrow l}$ that label the partial decay width of
a particle $k$ in any final state containing the particle $l$;
in our discussion they play a role just at late times when NLSPs  
decay into LSPs, giving the scaling we have used in Eq.~(\ref{eqn:rhophi}).
The first two terms refer, respectively, to processes of the
kind $(a)$ and $(b)$ in Eq.~(\ref{eq:list}), including all
possible SM final and initial states.  The symbol
$\langle\sigma_{ab} v_{ab}\rangle$ indicates a thermal average
of the cross section $\sigma_{ab} v_{ab}$; i.e.,
\begin{equation} 
\langle\sigma_{ab} v_{ab}\rangle =
\frac{1}{n_a^{\rm eq}\, n_b^{\rm eq}}
\int d^3p_a d^3p_b  f_a^{\rm eq}(E_a) f_b^{\rm eq}(E_b) \sigma_{ab} v_{ab}\,.
\end{equation}
As is evident from the form we wrote the Boltzmann equation,
interaction rates have to be compared with the expansion rate
$H$ of the Universe. In general, over a large range of
intermediate temperatures, the $\langle\sigma_{iX \rightarrow j}
v_{iX \rightarrow j}\rangle\, n_X^{\rm eq}$ terms will be larger
than the $\langle\sigma_{ij} v_{ij}\rangle \, n_j^{\rm eq}$ terms,
since we expect the cross sections to be of the same order in the
two cases, but the scattering rates will be more efficient as
long as they involve light background particles $X$ with
relativistic equilibrium densities $n_X^{\rm eq}$ that are much
larger than the nonrelativistic Maxwell-Boltzmann--suppressed
equilibrium densities  $n_j^{\rm eq}$ for the more massive particles
$j$. This implies that collision processes go out of equilibrium
at a smaller temperature, or later time, than pair-annihilation
processes. Writing explicitly the expression for 
$d/dx(Y_i/Y_i^{\rm eq}-Y_k/Y_k^{\rm eq})$, in which
Maxwell-Boltzmann--suppressed terms and terms in mass splitting
over mass scale can be neglected, one finds explicitly that in
the limit that collisional rates are much larger than the expansion
rate, for any $i$ and $k$, $Y_i(x)/Y_i^{\rm eq}(x)=Y_k(x)/Y_k^{\rm eq}(x)$, or 
equivalently,
\begin{equation}
Y_i (x) = \frac{Y (x)}{Y^{\rm eq}(x)} Y_i^{\rm eq}(x)\;,
\label{eq:sumratio}
\end{equation}
with $Y (x) \equiv \sum_k Y_k(x)$ and $Y^{\rm eq}(x) \equiv \sum_k Y_k^{\rm eq}(x)$.
At the temperature $T_{cfo}$, when 
$\sum_X \langle\sigma_{iX \rightarrow k} v_{iX \rightarrow k}\rangle\, n_X^{\rm eq}
\simeq H$, collision processes decouple and the relative number densities
become frozen to about
\begin{equation}
\frac{n_i(T)}{n_k(T)} 
= \frac{n_i^{\rm eq}\left(T_{cfo}\right)}{n_k^{\rm eq}\left(T_{cfo}\right)}
\simeq \frac{g_i}{g_k} \left(\frac{m_i}{m_k}\right)^{3/2} 
\exp\left(\frac{m_k-m_i}{T_{cfo}}\right)
,
\label{eq:ratio}
\end{equation}
up to the time (temperature) at which heavier particles decay into lighter 
ones.

The sum $Y(x)$ of the number densities has instead decoupled
long before.  Eq.~(\ref{eq:ratio})  is the relation that is
implemented to find the usual Boltzmann
equation~\cite{Griest:1990kh,edsjogondolo} for the sum over
number densities of all species compared to the sum of
equilibrium number densities, and that
shows that the decoupling for $Y$ occurs when the total effective annihilation 
rate becomes smaller than the expansion rate, at a temperature 
$T_{afo}$ that, as mentioned above, is much larger than $T_{cfo}$.  

To get an estimate for $T_{cfo}$, we can take, whenever a
channel is kinematically allowed, the (very) rough s-wave limit,
\begin{equation}
\sigma_{i \rightarrow k} v_{i \rightarrow k} \sim
\sigma_{ik} v_{ik} \sim 
\langle\sigma_{ik} v_{ik} \rangle \sim 
\frac{3 \cdot 10^{-27} \,{\rm cm}^3\,{\rm s}^{-1}}{ \Omega_\chi h^2}\;,
\end{equation}
where an approximate relation between annihilation rate and relic density has 
been used~\cite{Jungman:1995df}. There are then two possibilities depending on whether 
{\sl (i)} the background particle $X$ enforcing  collisional
equilibrium has a mass much 
larger than the mass splitting $\Delta m$ between the SUSY particles involved, 
or {\sl (ii)} the opposite regime holds. In the first case we find that 
$T_{cfo} \sim m_{X} / (10 -15)$, roughly the temperature at which $X$ itself 
(except for neutrinos) decouples from equilibrium.  From
Eq.~(\ref{eq:ratio}) we find that $n_i/n_k \simeq g_i/g_k$; i.e.,
they have comparable abundances.  In the opposite case, we find
instead that $T_{cfo} \sim \Delta m / (10-15)$ and hence
$n_i/n_k \simeq g_i/g_k \left(\frac{m_i}{m_k}\right)^{3/2}
\exp\left[(10-15) {\rm sign}(m_k-m_i)\right]$; i.e., the
abundance of the heavier particle is totally negligible compared
to the lighter one.

Long-lived stau NLSPs are kept in collisional equilibrium with
neutralinos by scattering on background $\tau^{\pm}$ and
emission of a photon. In this case we are clearly in the limit
{\sl (i)}, as $m_{\stau}-m_{\neut} \ll m_{\tau}$.  Since the
number of internal degrees of freedom for both staus and
neutralinos is 2, we find that the fraction of charged dark
matter in this model is
\begin{equation}
f_\phi=\frac{g_{\stau}}{g_{\neut}+g_{\stau}} = \frac{1}{2}\, .
\end{equation} 
More carefully, though, this is not a strict prediction of our
MSSM setup for a stau NLDP, and should be interpreted just as an
upper limit. In fact, if we have other SUSY particles that are
quasi-degenerate in mass with the LSP, and if they then
coannihilate and decouple from the neutralino at a later time than
the stau, either through mode {\sl (i)} and then
immediately decaying into neutralinos (in all explicit examples,
the decay into staus is strongly suppressed compared to the
decay into neutralinos), or through mode {\sl (ii)}, 
then $f_\phi$ is reduced to
\begin{equation}
f_\phi=\frac{g_{\stau}}{\sum_{i=1}^n g_i}.
\label{eq:smallerfrac}
\end{equation}
where the sum in the denominator involves the neutralino, the
stau and all SUSY particles with $T_{cfo}$ lower than the
$T_{cfo}$ for staus.  

In particular, if the lightest neutralino
is a nearly-pure higgsino, then the next-to-lightest neutralino
will also be a higgsino very nearly degenerate in mass, and the
lightest chargino will also be nearly degenerate in mass, with
mass splittings possibly smaller than $m_{\tau}$. 
In this case, charginos and neutralinos will be kept in collisional 
equilibrium through scatterings on $(\nu_l,l)$ pairs. At the same 
time, the collisional decoupling of staus might be slightly delayed
because of chargino-stau conversions through the emission 
of a photon and absorption of a tau neutrino; however this 
second process has a Yukawa suppression (as we are
considering Higgsino-like charginos) compared to the first, and
hence it is still guaranteed that the stau decoupling temperature is
larger than the chargino $T_{cfo}$ temperature. Since the mass splitting
between neutralino and chargino and that between lightest neutralino
and next-to-lightest neutralino cannot be smaller than few tens of MeV 
(due to loop corrections to the masses), decoupling will always happen in 
mode {\sl (ii)} and we do not have to worry about possible stau
production in their decays. Since $g_{\chi^+}=4$, applying the formula in 
Eq.~(\ref{eq:smallerfrac}) we find $f_\phi=1/5$. In the case of a wino-like lightest neutralino, instead, the only extra coannihilating partner would be the lightest chargino, yielding $f_\phi=1/4$. Finally, adding to this picture, e.g.,
a quasi-degenerate smuon and selectron, $f_\phi=1/7$ could be obtained.

\section{Long lived stau NLSPs in sample minimal models}
\label{sec:msugra}

We provide a few examples of well motivated theoretical
scenarios where neutralino-stau degeneracy occurs, possibly in connection
with further coannihilating partners driving low values of $f_\phi$. 
In surveying the possible models, the criterion we take here is that 
of the {\em composition} of the lightest
neutralino in terms of its dominating gauge eigenstate components. We thus
outline {\em bino-}, {\em higgsino-} and {\em wino-}like lightest-neutralino
benchmark scenarios. 

\subsection{A case with $f_{\phi}=1/2$: Binos in the mSUGRA model}

\begin{figure}
\centerline{\psfig{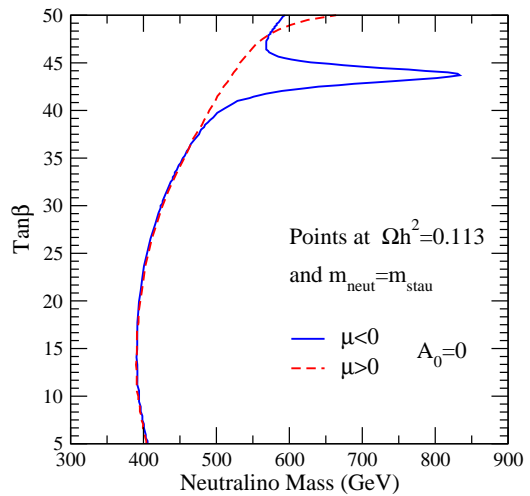}}
\caption{Points of the mSUGRA parameter space, at null trilinear
scalar coupling $A_0$, featuring $\Omega_{\neut}=0.11$ and
$m_{\neut}=m_{\stau}$, for both negative (solid line) and
positive (dashed line) sign of the $\mu$ parameter.}
\label{fig:PS}
\end{figure}

As we mentioned above, in the framework of
minimal supergravity (mSUGRA)~\cite{msugra}  one of the few cosmologically
allowed regions of parameter space is the tail where the neutralino and
the stau are quasi-degenerate.  In this case, coannihilations
reduce the exceedingly large bino-like neutralino
relic abundance to cosmologically acceptable values for
neutralino masses up to around 600 GeV. Coannihilation effects
depend on the relative mass splitting between the two
coannihilating species. Requiring a mass splitting as small as
those found above amounts, as far as the neutralino relic
density is concerned, to effectively setting
$m_{\neut}=m_{\stau}$. This, in turns, sets the mass of the
neutralino-stau system once a particular value of the relic
abundance is required. We plot in Fig.~\ref{fig:PS} points
fulfilling at once $m_{\neut}=m_{\stau}$ and
$\Omega_{\neut}=\Omega_{\rm CDM}\simeq 0.113$, the latter being
the central value as determined from the analysis of CMB data
\cite{CMBconstraints}. Result are shown
in the ($m_{\neut}$-$\tan\beta$) plane, with $\tan\beta$
the ratio of the vacuum expectation values of the two neutral 
components of the $SU(2)$ Higgs doublets, and at a fixed value of the
trilinear coupling $A_0=0$ (this latter quantity is, however,
not crucial here). The solid line corresponds to negative values
of the Higgsino mass parameter $\mu$, while the dashed line corresponds to positive values of $\mu$.  Notice that at $\mu<0$ the accidental overlap of the
heavy Higgs resonance with the coannihilation strip, around
$\tan\beta=43$, shifts the neutralino masses to larger
values. We point out that the two requirements of mass
degeneracy and of the correct relic density determine, at a
given value of $\tan\beta$, the required mass of the
neutralino-stau system, thus solving the residual mSUGRA
parameter space degeneracy, and making the present framework testable
and predictive. 

In the minimal configuration we have considered, only the lightest neutralino
and the lightest stau are playing a role, hence $f_{\phi}=1/2$. However,
since the mass splitting between the lightest stau and lightest smuon and selectron
is rather small, assuming a slight departure from universality in the scalar 
sector, two additional quasi-degenerate scalar particles can be obtained and 
the fraction of charged dark matter reduced to $f_{\phi}=1/4$.

\subsection{A case with $f_{\phi}=1/5$: Higgsino-like neutralinos}
\label{sec:higgsino}

\begin{figure}
\centerline{\epsfig{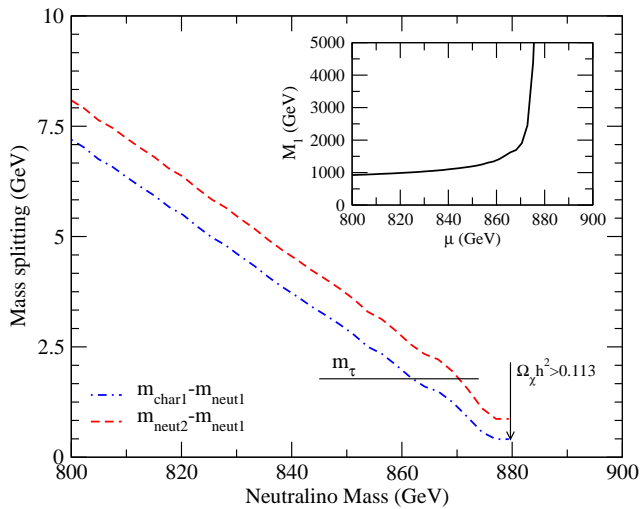}}
\caption{A low-energy parameterization of the higgsino-like neutralino case, at $\tan\beta=50$. 
In the smaller frame we indicate the points, on the $(\mu,M_1)$ plane, that
give $\Omega h^2\simeq 0.113$. The corresponding models are reproduced in the
larger frame, where we plot the relevant mass splittings of the chargino-neutralino system.}
\label{fig:higgsino}
\end{figure}

When the $\mu$ term is lighter than the gaugino masses $M_1$ and $M_2$, 
the lightest neutralino gets dominated by the higgsino component. 
This situation occurs, again within the mSUGRA model, in the so-called 
hyperbolic branch/focus point (HB/FP) region \cite{Chan:1997bi,Feng:1999zg}, where large values of the
common soft breaking scalar mass $m_0$ drive $\mu$ to low values. 
In this region, scalars are naturally heavy, at least in the minimal setup; however,
the occurrence of non-universalities in the scalar sector \cite{Profumo:2003em} may 
significantly affect the sfermion mass pattern. In particular, in a SUSY-GUT
scenario, soft breaking sfermion masses get contributions from 
$D$-terms whenever the GUT 
gauge group is spontaneously broken with a reduction of rank \cite{Drees:1986vd}. 
Light staus may naturally occur, for instance when the weak hypercharge $D$-term dominates and features negative values. In this case the hierarchy between 
diagonal entries in the soft supersymmetry-breaking scalar mass matrices is
$m^2_{E}\ll m^2_{U,D,Q,L}$. The $m^2_L$ term may also be lowered in presence of additional $D$-terms originating from the breaking of further $U(1)$ symmetries.

The relic neutralino abundance is here fixed by the interplay of multiple
chargino-stau-neutralino coannihilations. In the limit of pure higgsino,
the dynamics of these processes fixes the value of $\mu$ yielding a 
given relic neutralino abundance. On the other hand, a mixing with the
bino component along the borders of the HB/FP region may entail 
a larger spread in the allowed mass range, affecting the $\neut$ higgsino fraction. We sketch the situation 
in Fig.~\ref{fig:higgsino}, where we resort, for computational ease, to a low-energy 
parameterization of the above outlined scenario. The smaller frame shows
the points on the $(\mu,M_1)$ plane that produce the required amount of relic
neutralinos. The larger frame reproduces the values of the chargino-neutralino
and neutralino--next-to-lightest-neutralino mass splitting; the lines end in the
pure-higgsino regime. Suitable models, in the present framework, must also 
fulfill the mass splitting requirement 
$m_{\widetilde \chi _2, \widetilde \chi^+_1}-m_{\neut}<m_\tau$. 
This enforces the allowed neutralino mass range, at $\tan\beta=50$, between 870 and 880 GeV. Had we lowered the value of $\tan\beta$, the corresponding $m_{\neut}$ range would only have shifted to masses just a few tens of GeV lighter. 

As we have already mentioned, since we are dealing with a case with two 
neutralinos, a chargino and a stau quasi-degenerate in mass, we find 
$f_{\phi}=1/5$. Again, a smuon and a selectron can be added to this to shift the 
charged particle fraction to $f_{\phi}=1/7$.

\subsection{A case with $f_{\phi}=1/4$: Wino-like neutralinos}
\label{sec:wino}

A benchmark case where the lightest neutralino is wino-like is instead provided
by the {\em minimal Anomaly mediated SUSY breaking} (mAMSB) scenario \cite{mamsb}. 
In this framework, tachyonic sfermion masses are cured by postulating 
a common scalar mass term, $m_0$. The lightest sfermion turns out 
to correspond  to the lightest stau $\stau$ again. 
The latter, at suitably low $m_0$ values, may be degenerate
in mass with the lightest neutralino. We performed a scan of the mAMSB 
parameter space, requiring $m_{\neut}\simeq m_{\stau}$, and found that 
the correct relic abundance requires the neutralino masses to lie in the range 
$1250\lesssim m_{\neut} \lesssim 1600$ GeV, with the lower bound holding for small values of $\tan\beta$, while the higher for larger ones. The chargino-neutralino mass splitting is always below the $\tau$ mass, since $\neut$ in mAMSB is always a very pure wino, and it has large masses. We finally remark that here, as in the case of higgsinos, the occurrence of stau coannihilations {\em raises} the neutralino relic abundance, contrary to the standard result with a bino-like LSP.

Since, in this case, we have one neutralino, a chargino and a stau 
quasi-degenerate in mass, we find $f_{\phi}=1/4$. 

\subsection{The Stau lifetime}

\begin{figure}
\centerline{\psfig{file=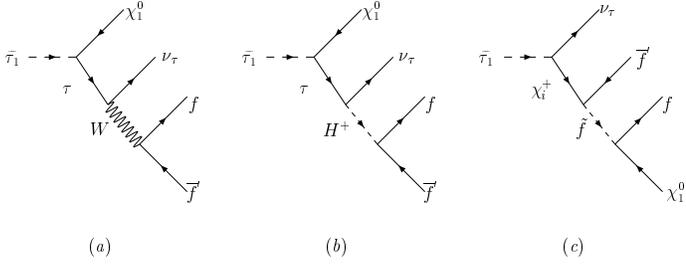,width=3.65in,angle=0}}
\caption{A few Feynman diagrams for four-body final states for
the process $\stau\,\, \rightarrow \,\,  \neut \, \nu_\tau \,
f\, \bar f^\prime,$. Diagram (a) is the dominant diagram;
diagrams of type (b) are sub-dominant, and diagrams 
of type (c) are sub-sub-dominant.}
\label{fig:feyn}
\end{figure}

The $\stau$ four-body decay proceeds through diagrams of the types sketched 
in Fig.~\ref{fig:feyn}. They come in three sets:
those with $W$ exchange, those with $H^\pm$ exchange, and those
with a sfermion exchange.  However, since $\Delta m\equiv
m_{\stau}-m_{\neut}$ is much
smaller than any supersymmetric-particle mass, the virtuality of
all diagrams except those featuring a $\tau$ exchange (diagrams
(a) and (b) in the Figure) is extremely large. Hence all diagrams but
those with a $\tau$ exchange will be suppressed by a
factor $(m_\tau/m_{\rm SUSY})^4\sim10^{-8}$, and the
interferences with the dominating diagrams by a
factor $(m_\tau/m_{\rm SUSY})^2\sim10^{-4}$. Of the two diagrams
with a $\tau$ exchange, however, the one with the $H^\pm$
exchange has a Yukawa suppressed $H^\pm f \bar f^\prime$ vertex,
which gives a suppression, with respect to the $W$-exchange
diagram, 
\begin{align}
\sim (m_\mu\tan\beta/m_W)^2(m_\tau\tan\beta/m_W)^2\sim 10^{-7} -
10^{-3}\nonumber\\ {\rm for}\ \ \tan\beta=5-50
\end{align}
in the most favorable muonic final
channel. Notice that the chirality structure of the couplings entails
that no interference between these two diagrams is present.
Diagrams with a charged Higgs or a sfermion exchange are
moreover further suppressed with respect to those with a $W$
exchange by a factor $(m_W/m_{\rm SUSY})^4$ which, depending on
the SUSY spectrum, can also be relevant.

In this respect, a very good approximation to the resulting stau
lifetime is obtained by considering only the first diagram,
whose squared amplitude reads
\begin{align}
|A|^2=\sum_{\rm final\, states}\,\frac{16(g_2/\sqrt{2})^4\, (p_{\nu_\tau}\cdot p_f) }{\left((p_\tau)^2-m_\tau^2\right)^2\left((p_W)^2-m_W^2\right)^2}\times \nonumber\\
\times \Big[|V_R|^2\left(2(p_{\neut}\cdot p_\tau)(p_\tau\cdot p_{\bar f^\prime})-(p_\tau)^2(p_{\neut}\cdot p_{\bar f^\prime})\right)+\nonumber\\
+m_\tau^2|V_L|^2(p_{\neut}\cdot p_{\bar f^\prime})-2m_{\neut}m_\tau{\rm Re}(V_L^*V_R)(p_\tau\cdot p_{\bar f^\prime})\Big],
\end{align}
with $V_{L,R}$ the left- and right-handed coupling of the
$\stau$ in the $\stau\tau\neut$ vertex. The sum is extended over
the final states of Eq.~(\ref{eq:fs}). For numerical purposes, 
the four-particles phase
space is integrated with the use of the Monte Carlo routine
Rambo, with final-state finite-mass corrections for all final
states.

\begin{figure}
\centerline{\psfig{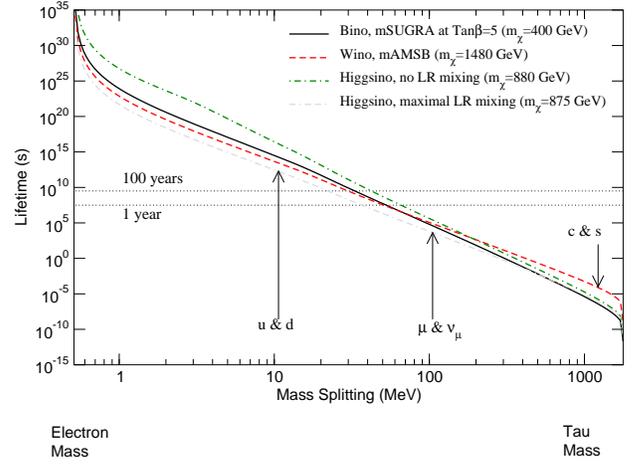}}
\caption{The stau lifetime, as a function of the mass splitting
with the lightest neutralino. The parameter space points are
defined by the two requirements $m_{\neut}\simeq
m_{\stau}$ and $\Omega_{\neut}=0.11$. For the wino and higgsino cases $\tan\beta=50$, while in all cases $A_0=0$ and $\mu>0$.}
\label{fig:staulife}  
\end{figure}

Fig.~\ref{fig:staulife} shows the stau lifetime for a sample of 
the supersymmetric scenarios outlined in the preceding sections.
We fully account for threshold effects in the
phase space, with the four contributions from electronic,
muonic, and first- and second-generation quarks. The quark masses
have been set to their central experimental values \cite{Hagiwara:2002fs}.
For the case of higgsinos, we reproduce the two extreme regimes when
the $m^2_L$ term is large (no Left-Right mixing) and when $m^2_L\simeq m^2_R$ (maximal Left-Right mixing). The differences in the lifetimes are traced back
to overall mass effects and to the values of the $V_L$ and $V_R$ couplings 
(for instance, $|V_L|\gg|V_R|$ in the wino case, while the opposite regime
holds for the case of higgsinos and no LR-mixing). 
In any case, we conclude that lifetimes of
the order of $1-100$ years are obtained with a mass splitting
$\Delta m=20- 70$ MeV.

\section{Dark Matter Searches and Collider Signatures}
\label{sec:colliders}

Unlike other charged long-lived NLDP scenarios, the
framework we outlined above has the merit of being,
in principle, detectable at dark-matter--detection and collider
experiments.

\begin{figure}
\centerline{\epsfig{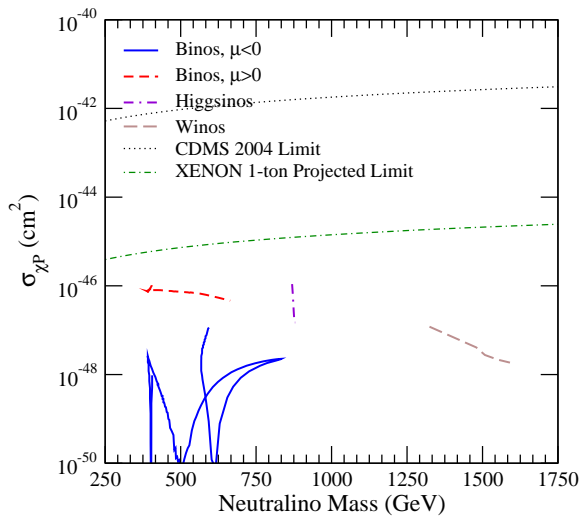}}
\caption{The spin-independent neutralino-proton scattering cross
section, as a function of the neutralino mass, along the
parameter-space points outlined in Fig.~\ref{fig:PS}. We also
indicate the current exclusion limits from the CDMS
experiment \cite{Akerib:2004fq}, and the expected reach
of the XENON 1-ton facility \cite{Aprile:2002ef}.}
\label{fig:SI}
\end{figure}

We show in Fig.~\ref{fig:SI} the spin-independent
neutralino-proton scattering cross section for the mSUGRA
parameter-space points of Fig.~\ref{fig:PS}, and for the higgsino and wino 
(mAMSB) cases, as discussed in sec.~\ref{sec:higgsino} and~\ref{sec:wino}, 
together with the current exclusion limits from the CDMS experiment
\cite{Akerib:2004fq}, and the future reach of the XENON
1-ton facility \cite{Aprile:2002ef}.  For binos, at $\mu>0$ most points lie
within less than one order of magnitude with respect to the future
projected sensitivity, making it conceivable that this scenario
may be tested in next-generation facilities. For negative $\mu$,
destructive interference among the lightest- and heavy-Higgs
contributions lead instead to cancellations in $\sigma_{\neut P}$. 
Finally, higgsino and wino detection rates respectively lie one and two orders of magnitude below the future expected sensitivity. 

Indirect-detection experiments look less
promising, even for winos and higgsinos, which always feature uncomfortably 
large neutralino masses.
We checked, for instance, that the expected muon flux
from the Sun, generated by neutralino pair annihilations, is at
most $10^{-(4-5)}$ muons per ${\rm km}^2$ per year, far below
the sensitivity of future neutrino telescopes like IceCube
\cite{icecube}.

Turning to collider experiments, considering as the searching tool the usual missing 
transverse-energy channels, dedicated studies have shown that the
mSUGRA coannihilation strip will be within LHC reach, mainly through
in a mass range that extends up to $m_{\neut}\lesssim 550$ GeV 
along the coannihilation strip, quite independently of $\tan\beta$
\cite{Baer:2003wx}. Concerning higgsinos and winos, instead, the relevant
mass range we study here appears to be beyond standard LHC searches \cite{Baer:2000bs}.

Even at high-energy colliders, the peculiar and distinctive feature of this 
scenario is however represented by the long-lived stau.
The production of what have sometimes been 
dubbed long-lived {\em charged massive particles}
(CHAMP's) \cite{Acosta:2002ju} has been repeatedly addressed.
Exclusion limits were determined by the CDF Collaboration
\cite{Abe:1992vr,Acosta:2002ju}, and the future reach of the LHC
and of a future Linear Collider has been also assessed for this
broad class of exotic new particles
\cite{Nisati:1997gb,Mercadante:2000hw}. The case of a
stable\footnote{Here {\em stable} simply means that the decay
length is much larger than the detector size.} stau has been, in
particular, considered several times, since it occurs in the context
of various SUSY-breaking models, like gauge-mediated
SUSY-breaking (GMSB) scenarios \cite{Giudice:1998bp}; see also, e.g.,
Ref.~\cite{Feng:1997zr,Baer:2000pe}. We remark that, contrary to GMSB,
in the present framework the production of long lived staus at accelerators 
is expected to come along together with the production of neutralinos, 
thus making the two scenarios distinguishable, at least in principle. 
 
A long-lived stau would behave as a highly
penetrating particle, appearing in the tracking and muon
chambers of collider detectors with a small energy deposit in
calorimeters. Staus, depending on their velocities, would produce
either highly-ionizing tracks in the low-$\beta$ regime, or, if
quite relativistic, they would appear similar to energetic
muons. In this latter case, one could look at excesses of dimuon
or multi-lepton events as a result of superparticle production;
for example, considering the ratio
$\sigma(\mu^+\mu^-)/\sigma(e^+e^-)$. In case additional particles
have masses close to the neutralino-stau system, the total
production cross section of superparticles would be greatly
enhanced. Long-lived charginos may also give interesting accelerator
signals \cite{Feng:1999fu}.  

While the reach of the Tevatron appears to be insufficient to
probe the parameter space of the models we are considering here
\cite{Feng:1997zr}, the discovery of this kind of scenarios at
the LHC, though challenging, looks quite conceivable, particularly in the case 
of binos. In fact, even in the less promising case in
which the particle spectrum does not feature any particle close
in mass with the $\neut-\stau$ system, the highly-ionizing--track
channel should cover a mass range widely overlapping that
indicated in Fig.~\ref{fig:PS}.  On the other hand, excess
dimuon events could provide an independent confirmation,
although the 5-$\sigma$ LHC reach for CHAMP's in this channel alone has
been assessed to lie around 300 GeV
\cite{Feng:1997zr}. Moreover, if additional coannihilating
particles (charginos, smuons, or selectrons) are present, the
discovery at the LHC would certainly look even more promising
\cite{Baer:2000pe}.

A further recently proposed detection technique for long-lived
staus is represented by trapping these particles into large
water tanks placed outside the LHC detectors
\cite{Feng:2004yi}. Following the results of
Ref.~\cite{Feng:2004yi}, a 10-kton  water tank may be capable of
trapping more than 10 staus per year, if the mass
$m_{\stau}\simeq 400$ GeV.  The subsequent decays could then be
studied in a background-free environment. More than twice as many
sleptons would also get trapped in the LHC detectors, although
in this case a study of the stau properties would look more
challenging \cite{Feng:2004yi}.

\section{Conclusions}
\label{sec:conclusion}

We have examined a scenario in which a fraction $f_{\phi}$ of the
cold-dark-matter component in the Universe is generated in the
decay of long-lived charged particles and shown that 
a scale-dependent (or `running') spectral index is induced. The power spectrum on scales smaller than the horizon size when the age of the Universe is equal 
to the charged-particle lifetime gets suppressed by a factor $(1-f_{\phi})^2$. Such a feature might
be singled out unambiguously by future measurements of the power 
spectrum of neutral hydrogen through the 21-cm line, obtaining direct 
information on the charged particle lifetime and $f_{\phi}$.

On the contrary, current and future tests for departures from a scale-invariant power spectrum based on Lyman-$\alpha$ data may fail
to uniquely identify the scenario we propose. In fact, we have estimated the 
modifications to the non-linear power spectrum at the redshifts and 
wavenumbers currently probed by Lyman-$\alpha$ forest data, and 
shown that these resemble (but are different in detail) those in models
with a constant running of the spectral index $\alpha_{s}$.  We expect, based on this resemblance, models with $f_{\phi}$ in the range 
$1/2 - 1/7$ (as predicted in some explicit models we have 
constructed) to be compatible with current cosmological data for lifetimes in the range 
$1 - 20$~yr. We have also verified that constraints from the primordial light-element abundances and distortions to the CMB spectrum are not violated.

From the particle-physics point of view, we have shown that the proposed
scenario fits nicely in a picture in which the lightest neutralino in SUSY 
extensions of the standard model appears as the cold-dark-matter candidate, 
and a stau nearly degenerate in mass with the neutralino as the long-lived charged
counterpart. A small mass splitting
forces the stau to be quasi-stable, since the phase space allowed
in the its decay process gets sharply reduced. At the same time, it
implies that neutralino and stau are strongly linked in the process of
thermal decoupling, with the charged species playing the major role.  Owing to these
coannihilation effects, the current neutralino thermal relic abundance is 
compatible with the value inferred from cosmological observatations and, at early times, the stau 
thermal relic component is at the correct level.

We have described several explicit realizations of this idea in
minimal supersymmetric frameworks, including the minimal supergravity
scenario, namely the supersymmetric extension of the standard model with
smallest possible parameter space. We have pointed out that charged dark matter
fraction from 1/2 to 1/7 (or even lower) can be obtained and that stau
lifetimes larger than 1~yr are feasible. We have also shown that some of the
models we have considered may be detected in future WIMP direct searches, and
discussed the prospects of testing the most dramatic feature of the model
we propose, i.e. the production of long-lived staus at future high energy particle
colliders.

\begin{acknowledgments}
We thank R. R. Caldwell and U. Seljak for useful discussions.  SP acknowledges fruitful help and elucidations from Y. Mambrini about the SDECAY package.
KS acknowledges the support of a Canadian NSERC Postgraduate
Scholarship.  This work was supported in part by NASA NAG5-9821
and DoE DE-FG03-92-ER40701.
\end{acknowledgments}

\end{document}